\documentclass[aps,pre,reprint,showpacs,superscriptaddress
]{revtex4-1} 
\usepackage{graphicx}
\usepackage{amsfonts}
\usepackage{amsmath}
\usepackage{amssymb}
\usepackage{txfonts} 
\usepackage{times}
\usepackage{subfigure}
\usepackage{textcomp}
\usepackage{color}
\definecolor{gray}{gray}{0.8}
\usepackage{dcolumn}
\usepackage{bm}
\usepackage{hyperref}

\newcommand{\beq}{\begin{equation}}     \newcommand{\eeq}{\end{equation}}
\newcommand{\beqa}{\begin{eqnarray}}    \newcommand{\eeqa}{\end{eqnarray}}
\newcommand{\bde}{\begin{description}}  \newcommand{\ede}{\end{description}}
\newcommand{\ben}{\begin{enumerate}}    \newcommand{\een}{\end{enumerate}}
\newcommand{\ind}{\indent }             
\newcommand{\noi}{\noindent\mbox{}}

\newcommand{\la}{\langle}               \newcommand{\ra}{\rangle}

\newcommand{\intII}{{\int^{+\infty}_{-\infty}}}

\newcommand{\eqn}[1]{\beq{ #1 }\eeq}

\newcommand{\inv}[1]{{\frac{1}{#1}}}

\newcommand{\inRbracket}[1]{{\left({#1}\right)}}
\newcommand{\inSbracket}[1]{{\left[{#1}\right]}}

\newcommand{\inAverage}[1]{{\left\la{#1}\right\ra}}

\newcommand{\meq}{m^{\rm (eq)}}

\begin{document}
\title{ Progressive Quenching - Globally Coupled Model}

\author{ Bruno Vent\'ejou} \affiliation{Gulliver, CNRS-UMR7083, ESPCI, 75231 Paris, France} \author{Ken Sekimoto} \email{ken.sekimoto@espci.fr}\affiliation{Mati\`{e}res et Syst\`{e}mes Complexes, CNRS-UMR7057, Universit\'e   Paris-Diderot, 75205 Paris, France}  \affiliation{Gulliver, CNRS-UMR7083, ESPCI, 75231 Paris, France} 
\begin{abstract}
We study the processes in which fluctuating elements of a system are progressively fixed (quenched) while keeping the interaction with the remaining unfixed elements. If the interaction is global among  Ising spin elements and if the unfixed part is re-equilibrated each time after fixing an element, the evolution of a large system is martingale about the equilibrium spin value of the unfixed spins. 
Due to this property the system starting from the critical point yields the final magnetization 
whose distribution shows non-Gaussian and slow transient {}{ behavior} 
with the system size.
\end{abstract}
\maketitle

\section{Introduction} 
{}{Since the end of the last century much development has been made in the physics of stochastic processes out of equilibrium of a finite system interacting with a heat bath (or baths) and under the influences of an external system (or systems).
There the focus has been mostly on the cases in which the division between the system and the external system is fixed. In real world, however,
} 
we sometimes encounter the situations in which system's degrees of freedom become progressively fixed. 
When a molten material as a fluid system is pulled out from a furnace and is quickly cooled down \cite{Damien2017}, the fluid degrees of freedom associated to fluid particles are progressively fixed (quenched).
{}{Although analogy is not close, we might also} consider the process of decision-making by a community, in which 
 each member progressively makes up her or his mind before the referendum. 
In both examples, the already fixed part can influence the behavior of the part whose degrees of freedom are not yet fixed. 
It is largely unknown what types of generic aspects are in this type of problems, {}{and our object is to find them out. We
propose to name this problem  the ``progressive quenching.}''

A prototype of this problem has been studied long time ago in the context of the phason fluctuations of quasi-crystal \cite{phason-freezing-PhA}. The phason is a Goldstone mode of the quasicrystalline order, whose evolution is modelled by a diffusive dynamics under non-conservative thermal noise. If we simplify the problem to 1D, a scalar phason field, $\psi(x,t),$ obeys $\partial\psi/\partial t=D\partial^2 \psi/\partial x^2+\xi(x,t),$ where $\xi(x,t)$ is a Gaussian white noise uncorrelated both in space ($x$) and in time ($t$). In equilibrium the spatial correlation is known to obey $\la |\psi(x+r,t)-\psi(x,t)|^2\ra \sim |r|.$  The progressive quenching fixes the value of $\psi(x,t)$ at the front position, $x=V t,$ which moves in $+x$ direction at a constant speed $V(>\! 0).$ Then the spatial correlation in the fixed part shows the different statistics $\la |\psi(x+r)-\psi(x)|^2\ra \sim |r|^{3/2}/\ell_D^{1/2},$ over the length-scale inferior to the diffusion length, $\ell_D\equiv D/V.$
Similar kind of study can be done for 1D spin models \cite{PQ-chain}. 
In the above examples the quenched part acted as an external field but it was applied only in the vicinity of the quenching front. The interest there was whether or how the progressive quenching modifies the spatial statistics of the system's configuration with respect to the equilibrium one.
In the present {}{Letter}, we will focus on a complementary case, where the quenched part influences 
the whole unquenched part of the system.
In the context of the decision-making, the results of preliminary survey {}{updated frequently (e.g. on the internet)} before the referendum {}{will represent those who already made up their mind and they} can influence all those people who do not yet make up their mind.
As a simple and concrete model, we take up the globally coupled, or infinite-range, Ising model, with the weak coupling in the terminology of \cite{Hilfer-2003}, and adopt the stepwise re-equilibration of the unfixed part of spins detailed below. The most interesting case is when the system is initially at the critical point. 
(Often the important referendums are done when the public opinion is little stable.)
Our main finding is that if we regard the mean equilibrium spin in the unfixed part as the stochastic process along the number of fixed spins as "time", the process shows the approximate or asymptotic martingale property  {}{with respect to the sequentially fixed spins whether or not the process starts from the critical point. In general we say a discrete stochastic process $\{X_T\}$ ($T=0,1,\ldots$) is martingale with respect to the stochastic process $\{Y_1,\ldots,Y_T\}$ if the conditional expectation of the former satisfies $E[X_{T+1}|\{Y_1,\ldots,Y_T\}]=X_T$ and $X_T$ is determined as function of $\{Y_1,\ldots,Y_T\}.$ 
In the present context $X_T$ is the mean re-equilibrated spin after $T$ of spins have been fixed, and $\{Y_1,\ldots,Y_T\}$ stands for the history of fixed spins  \cite{martingale-book}.
While the martingale properties have been used in physics as technical tool, its physical meanings and consequences have been rarely exploited. It is only very recently that \cite{martingale-Gupta2011,Roldan-prX2017} recognized the detailed fluctuation theorems as the martingale property of the path probability ratios. 
Our present study uncovers a new physical mechanism of the martingale property, whose important consequence is that the initial stochastic history has strong and long-lasting effects on the later process 
\footnote{In analogy to the referendum, a tentative interpretation is that the opinion of the first few determined persons has often a decisive impact.}.} 

\section{Setup of problem}
\subsection{System} 
\begin{figure}[t!!]
\centering
{\includegraphics[width=2.5cm,angle=-90.]{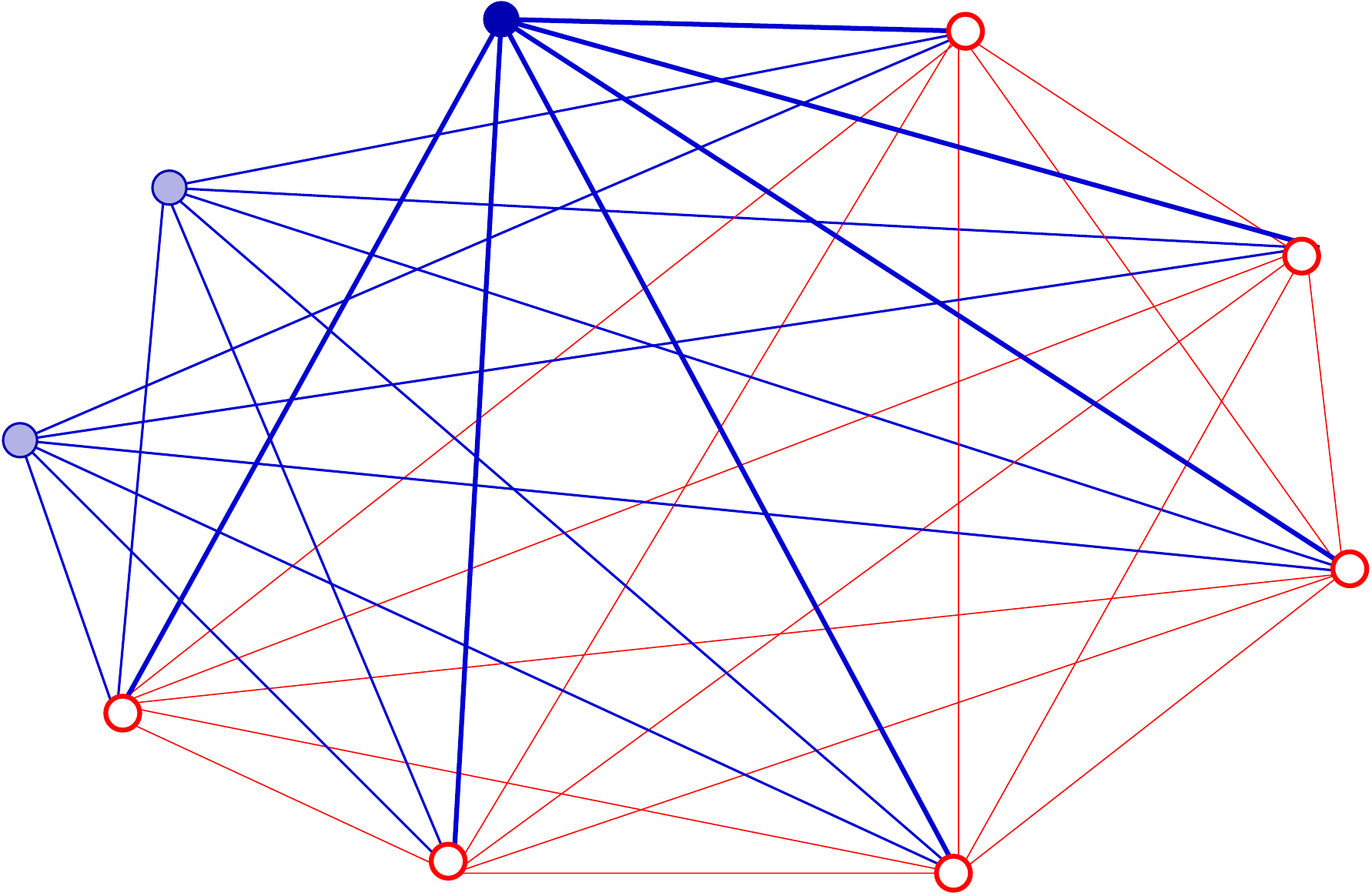}}  
\caption{ Among $N_0=9$ spins (vertices) the three {}{($T=3$)} spins {}{on} the blue (filled) dots  have been fixed. These spins apply a field  $h$ on the remaining six {}{($N=6$)} spins {}{on} the red (open) circles, which obey canonical equilibrium.
Each edge denotes the coupling, $j_0/N_0,$ and those coupling among the fixed spins have been omitted.
\label{fig:schematic} 
}
\end{figure}

 We consider the ferromagnetic Ising model on a complete network, that is, the model in which any one of spins interacts with all the other spins with equal coupling constant, {}{$j_0/N_0,$ where $N_0$ is the total number of spins.} See Fig.~\ref{fig:schematic}.
 {}{When $N(\le N_0)$ spins are unfixed and in canonical equilibrium under a field $h$, we use 
the energy function,  }
 \eqn{
 H_{N,h}=- \frac{j_0}{N_0} \sum_{1\le i<j\le N}s_i s_j - h\sum_{1\le i\le N} s_i,
 }
where each spin takes the values $\pm 1,$ {}{and  $m^{\rm (eq)}_{N,h}=$ $\la s_i\ra^{\rm (eq)}$ $\equiv  N^{-1}\partial [\ln \sum_{\{s_1,\ldots,s_N\}}e^{-\beta H_{N,h}}]/\partial h$ gives the mean equilibrium spin.}
{ 
Unless noticed explicitly all the calculations of $m^{\rm (eq)}_{N,h}$ and its derivatives with respect to $N$ or $h$ are done using the Hubbard-Stratonovich transformation applied to the canonical partition function, that is, 
 \eqn{\label{eq:IntegZ}
\sum_{\{s_1,\ldots,s_N\}}e^{-\beta H_{N,h}}=c_{N}\intII dx e^{-\frac{\beta j_0 x^2}{2N_0}}
\inSbracket{2\cosh\inRbracket{\frac{\beta j_0}{N_0}x+\beta h}}^N,}
 where $c_{N}$ is a number independent of $h.$ We avoided principally the usage of the saddle-point evaluation, since such approximation brought non-negligible differences in the system-size {}{dependence}   discussed later. We recall, nevertheless, that for $N=N_0\to\infty$ the Curie point is $j_0=1$ }
{}{because the mean equilibrium spin, $m^{\rm (eq)}_{N=N_0,h}$ obeys $m^{\rm (eq)}_{N=N_0,h}=\tanh(\beta[ j_0 m^{\rm (eq)}_{N=N_0,h}+h])$ in this limit.
The progressive quenching proceeds under fixed values of the coupling strength $j_0/N_0$ and the inverse temperature $\beta$.  Hereafter, we will write $\beta j_0$ and $\beta h$ as $j_0$ and $h,$ respectively.}

\subsection{Protocol} 
We start with all the $N_0$ spins in equilibrium with zero external field, $h=0.$
We fix the first spin, $s_1,$  either at $+1$ or at $-1$ with equal probabilities. 
Once it done, we let re-equilibrate the remaining $N_0-1$ spins before fixing the second spin, $s_2.$
When the $T(>0)$ spins, $\{s_1,\ldots,s_T\},$ have already been frozen, the magnetization of these spins is $M_T\equiv \sum_{i=1}^T s_i.$ We then let re-equilibrate those $N_0-T$ unfixed spins under the magnetic field $h=h_T$ which is  induced by the fixed magnetization, that is, $h_T=(j_0/N_0)M_T.$ 
The equilibrium spin value of the unfixed spins at that stage is $m^{\rm (eq)}_{N_0-T,{}h_T},$  where $N=N_0-T.$
Then we fix the $(T+1)$-th spin, $s_{T+1},$ at the state where it took at that moment:   $s_{T+1}$ takes the value $\pm 1$ with the probabilities, respectively, 
\eqn{\label{eq:proba}
\mbox{Prob}(s_{T+1}=\pm 1)=\frac{1\pm m^{\rm (eq)}_{N_0-T,{}h_T}}{2}.
}
We repeat this operation until all the $N_0$ spins are fixed.
\subsection{Biased random walk} 
\begin{figure}[t!!]
\centering
\subfigure[\null]
{\label{fig:TN-schema} 
\includegraphics[width=2.cm,angle=-0.]{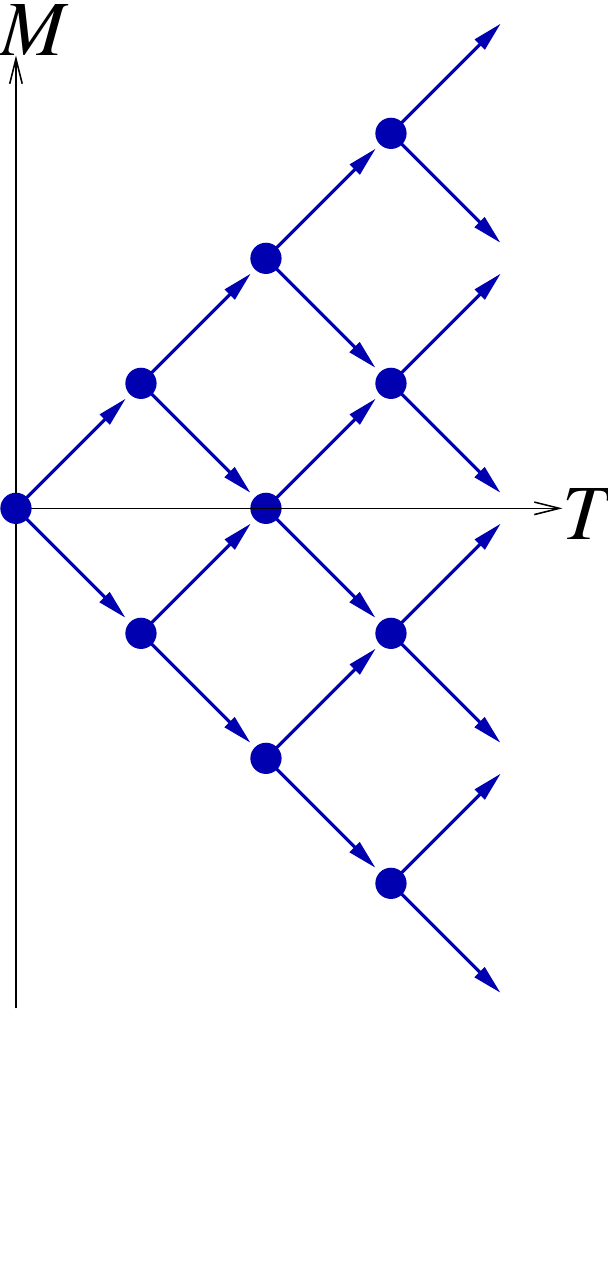}
}  
\hspace{5mm}
\subfigure[\null]
{\label{fig:off-critical-samples}    
\includegraphics[width=3.5cm,angle=-0.]{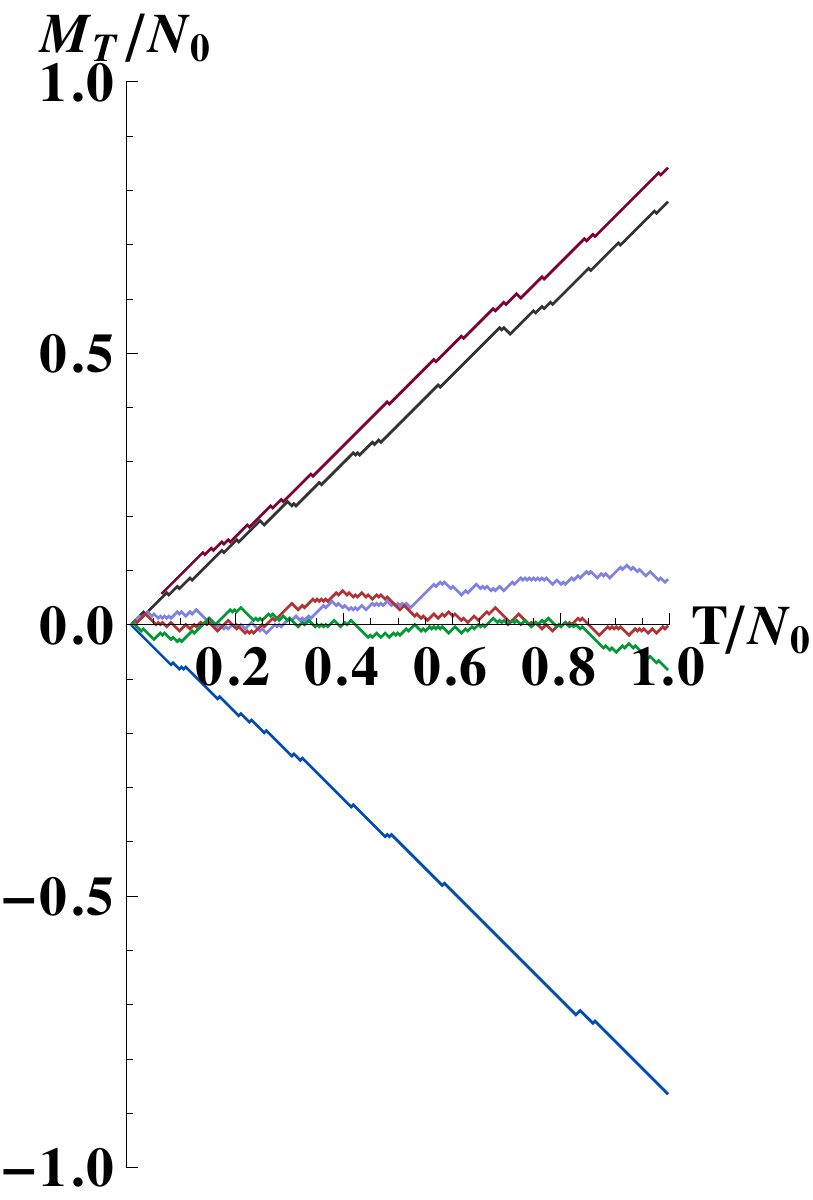}
}  
\hspace{-5mm}
\subfigure[\null]
{\label{fig:m-contour}    
\includegraphics[width=7cm,angle=90.0]{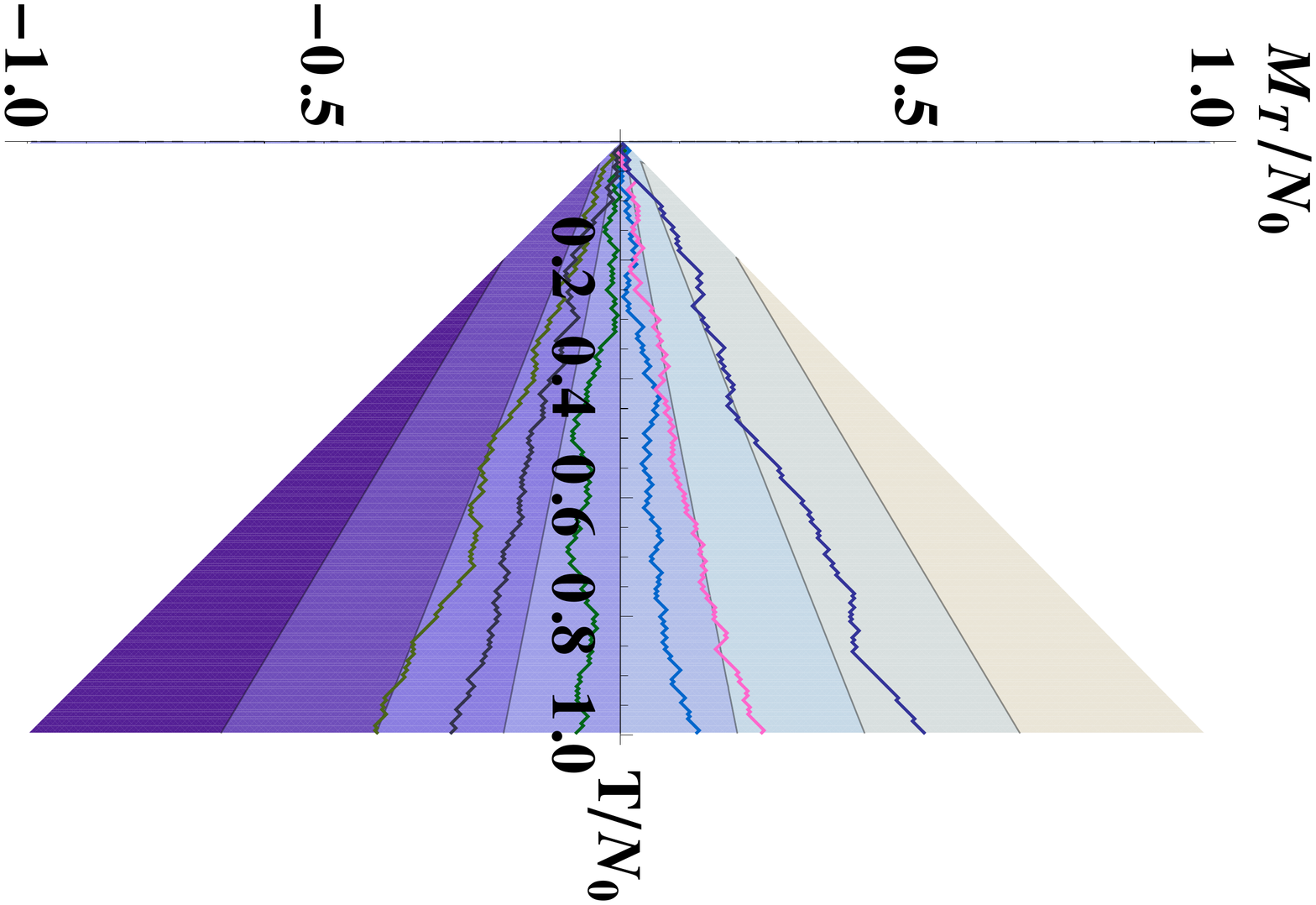}
}  
\caption{
 \label{fig:TN-samples} 
(a) 
 Markovian biased random walk corresponding to the present model of progressive quenching. From each node (blue (thick) dot)  in the transition network, $(T,M),$ the possible branched transitions, $M\to M\pm 1,$ occur with the probabilities, $(1\pm m^{\rm (eq)}_{N_0-T,{} (j_0/N_0)M})/2,$ which corresponds, respectively, to fixing $s_{T+1}$ at $\pm 1.$ 
(b) Three sample histories with $j_0=1.5$ (curves near the diagonals), and three others with $j_0=0$ (curves near the horizontal axis) are shown by different colors (brightness) for the system with the total size $N_0=256.$
(c) 
The six sample histories (curves of different colors (brightness)) with $j_0=j_{0,c}(\simeq {1.030}),$ the ``critical coupling''
with the size $N_0=256,$ are superposed 
on the contour plot of $m^{\rm (eq)}_{N_0-T,{}h_T}$ for the same $j_0$ 
(almost straight lines inside the triangle with gradient of color (brightness)).
 The value of $m^{\rm (eq)}_{N_0-T,{}h_T}$ is positive [negative], respectively, above [below] the horizontal axis.
}
\end{figure}

The above model defines the discrete-time markovian biased random walk for which the ``time'' is the number of fixed spins, $T$, and the ``position'' is the magnetization of fixed spins, $M,$ see Fig.~\ref{fig:TN-schema}.   
\noi The discrete stochastic evolution of $M_T$ is given as
\eqn{\label{eq:langevin}
M_{T+1}= M_{T}+s_{T+1}
}
for $0\le T\le N_0-1$ and $M_0=0.$ The random variable $s_{T+1}$ obeys the probabilities given in (\ref{eq:proba}) with $h_T=(j_0/N_0)M_T$ .
{}{
Then the path probability $\mbox{Prob}(\{s_1,\ldots,s_{T}\})$
for the history of quenching spins up to the step $T$, $\{s_1,\ldots,s_{T}\},$ can be constructed and is written as 
\beq \label{eq:pathprob}
\mbox{Prob}(\{s_1,\ldots,s_{T}\})=\inv{2^T}\prod_{T'=1}^{T}\inRbracket{1+s_{T'}m^{\rm (eq)}_{N_0-T'+1,{}\,h_{T'-1}}}.
\eeq
If we notice that the factor $2^{-T}$ is the path probability of {\it unbiased} quenching of spins, the product $R_T\equiv \prod_{T'=1}^{T}\inRbracket{1+s_{T'}m^{\rm (eq)}_{N_0-T'+1,{}\,h_{T'-1}}}
$ is the so-called the Radon-Nikodym derivative [functional] relating the biased and unbiased processes
\cite{Baxter-Rennie1996}, and $R_T$ is 
martingale with respect to this unbiased process, that is, 
$E_0[R_{T+1}|\{s_1,\ldots,s_T\}]=R_T,$ where $E_0$ means to take the (conditional) expectation of over the { unbiased} process. This is essentially the viewpoint in which 
\cite{martingale-Gupta2011,Roldan-prX2017} 
introduced a physical implication of the martingale in the equalities of Jarzynski \cite{Jarzynski97} and Crooks \cite{JeqCrooks98}, see Appendix.\ref{app:RN} for more explanations. 
In the progressive quenching process, however, we will show that the mean equilibrium spin, $m^{\rm (eq)}_{T}\equiv m^{\rm (eq)}_{N_0-T,h_T},$ shows asymptotically the martingale property by a different physical mechanism from the Radon-Nikodym derivative.}

\section{Results}
\subsection{Path samples}  
In Fig.~\ref{fig:off-critical-samples} we show representative sample histories, the three with $j_0=1.5$ (those curves near the diagonals) and the other three with $j_0=0$ (those near the horizontal axis). Both axes are normalized by the whole system size, $N_0$.
 In the former case ($j_0=1.5$) the system shows the typical ferromagnetic behavior; the initial fixed spin, $s_1$, induces a large magnetization in the unfixed part, which in turn biases the polarity of the spin to be fixed subsequently. 
In the latter case ($j_0=0$), where all the spins are unbiased and independent, the histories are the unbiased random walks. The final magnetization, $M_{T=N_0},$ then obeys the binomial probability distribution with zero mean and the variance, $N_0/4,$ and approaches the Gaussian distribution for $N_0\gg 1$ by the central limit  theorem.

Our interest is rather to understand what occurs between the two extreme cases mentioned above. Hereafter, we will focus on the system that starts from the ``critical'' point under zero external field ($h=0$). For finite system, $N_0<\infty,$ we define the ``critical'' point, $j_0=j_{0,\rm crit},$ through the extrapolation of the Curie law, $\chi^{-1}\propto j_{0,\rm crit}-j_0,$ from the paramagnetic side,
$j_{0,\rm crit}> j_0,$ with the susceptibility, $\chi=\partial m^{\rm (eq)}_{N_0,{}h}/\partial h|_{h=0}$ 
\footnote{The value of $j_{0,\rm crit}$ empirically fits well with $j_{0,\rm crit}-1 \simeq 5.06 \,{N_0}^{-0.933}$ over the range  $N_0= 2^5$-$2^{13}$.}.  
 Several representative histories of $M_T$ are shown in Fig.~\ref{fig:m-contour}. 
 We notice immediately that the curves are not like the unbiased random walk. Rather, $M_T$ in the late stages varies mostly linearly with $T.$ This feature is also common to the ``ferromagnetic'' case ($j_0=1.5$) in Fig.~\ref{fig:off-critical-samples}. In Fig.~\ref{fig:m-contour}
  we superposed the histories on the contour and grayscale (color gradient) map of the equilibrium spin, $m^{\rm (eq)}_{N_0-T,{}h_T}.$  
We there observe qualitatively that the individual histories like to keep the value of $m^{\rm (eq)}_{N_0-T,{}h_T}.$

\subsection{Martingale process in progressive quenching}\label{subsec:martingale}
Associated {}{to} the above observation we analytically found that, for $N_0\gg 1$ 
and $N\equiv N_0-T\sim N_0,$ the {\it stochastic process} $m^{\rm (eq)}_{T}\equiv m^{\rm (eq)}_{N_0-T,{}h_T}$ {\it vs} $T$ is approximately martingale with respect to $\{s_1,\ldots,s_T\}$,
that is, 
\beq \label{eq:meqISmartingale}
E[m^{\rm (eq)}_{T+1}|\{s_1,\ldots,s_T\}]=m^{\rm (eq)}_{T}+\mathcal{O}({N_0}^{-2}).
\eeq 
Eq.(\ref{eq:meqISmartingale}) explains why the numerical result of $M_T$ {\it vs} $T$ more or less follows the contours of the equilibrium spin, $m^{\rm (eq)}_{N_0-T,(j_0/N_0)M}=\mbox{const.}$
In fact, using (\ref{eq:proba}) we have 
\beqa \label{eq:mTexpansion}
&&
\!\!\!\!\! E[m^{\rm (eq)}_{T+1}|\{s_1,\ldots,s_T\}]
\cr
&&=\sum_{s_{T+1}\in \{-1,1\}}
\frac{1+s_{T+1} m^{\rm (eq)}_{T}}{2} m^{\rm (eq)}_{N_0-(T+1),{}\frac{j_0}{N_0}(M_T+s_{T+1})}
\cr
&&=
m^{\rm (eq)}_{T}
-{\frac{\partial m^{\rm (eq)}_{T}}{\partial N}}
+\frac{j_0}{2N_0}\frac{\partial {(m^{\rm (eq)}_{T})}^2}{\partial h_T}
+\mathcal{O}(\inv{{N_0}^2}),
\eeqa
{}{where  $m^{\rm (eq)}_{T}$ has been regarded as function of $N=N_0-T$  and $h_T=(j_0/N_0)M_T.$
Here the essential $N$-dependence is 
through the effective coupling constant, $j_T\equiv j_0(N/N_0),$ the parameter which appears when
the partition function for these spins, $\sum_{\{s_1,\ldots,s_N\}}e^{-\beta H_{N,h_T}},$
is rewritten as $\propto \sum_{\{s_1,\ldots,s_{N}\}}e^{N(j_T\frac{\overline{m}^2}{2}+h_T\overline{m})},$
with $\overline{m}\equiv (\sum_{i=1}^N s_i)/N$ being the mean unquenched spin.
We can show  that the second and the third terms on the r.h.s. of (\ref{eq:mTexpansion})
becomes $  ({j_0}/{2N_0}){\partial }
[{\la{ \overline{m}^2}\ra^{\rm (eq)}-({\inAverage{ \overline{m}}^{\rm (eq)}})^2}]/{\partial h}$
and, therefore,
cancel {with each other} to the order $\mathcal{O}({N_0}^{-1}).$  The derivation is given in Appendix.\ref{app:proof}, and we arrive at (\ref{eq:meqISmartingale}).

The error bound $\mathcal{O}({{N_0}^{-2}})$ in (\ref{eq:meqISmartingale}) is crucial:
When we iteratively apply this form backwards up to $T=1$ by fixing $M_1=s_1=1,$ there still holds $$E[m^{\rm (eq)}_{T+1}|s_1=1]=m^{\rm (eq)}_{N=N_0-1,h=j_0/N_0}+\mathcal{O}({{N_0}^{-1}}).$$ 
For example, for $N_0=2^8$, we numerically verified that the relative error to $m^{\rm (eq)}_{N=N_0-1,h=j_0/N_0} (\simeq 0.096)$
is of 0.2\% for $2^5<T<2^8.$ 
Later we will argue the physical mechanism of the cancellation in (\ref{eq:mTexpansion}).
 }
\subsection{Origin of {}quasi-straightness of equilibrium spin contours}\label{subsec:contour}
{}{In Fig.~\ref{fig:m-contour} we also notice that the contours of $m^{\rm (eq)}_{N_0-T,{}h}$ with $h=(j_0/N_0)M$  do not pass through the origin but are almost straight.
As function of $M$ and $T,$ the condition of the contour of $m^{\rm (eq)}_{N_0-T,{}h}$ 
reads,
\beq
0=dm^{\rm (eq)}_{N_0-T,{}h}
=\frac{j_0}{N_0}\frac{\partial m^{\rm (eq)}_{N_0-T,{}h}}{\partial h} dM+\frac{\partial m^{\rm (eq)}_{N_0-T,{}h}}{\partial T}dT.
\eeq

Using the above mentioned cancellation of terms in (\ref{eq:mTexpansion}), we find 
\eqn{\label{eq:dMdT}
\left.\frac{dM}{dT}\right. 
= -\,\inRbracket{\frac{\partial m^{\rm (eq)}_{N_0-T,h}}{\partial h}}^{-1}\,\frac{\partial m^{\rm (eq)}_{N_0-T,h}}{\partial T}= m^{\rm (eq)}_{N_0-T,h}+\mathcal{O}({N_0}^{-1})}
along the contour, $m^{\rm (eq)}_{N_0-T,h}=\mbox{const.}$
Moreover, (\ref{eq:dMdT}) tells that the mean tangent of each trajectory approximately gives the value of the mean equilibrium spin, $m^{\rm (eq)}_{N_0-T,h},$ memorized by the martingale property.}
\subsection{Compensation mechanism behind martingale property}  
The origin of the martingale property of the mean unfixed spin, $m^{\rm (eq)}_{T}\equiv m^{\rm (eq)}_{N_0-T,{}(j_0/N_0)M_T},$ 
is the compensation between the increment of the quenched field, $h_T=(j_0/N_0)M_T,$ 
and the decrease in the effective coupling parameters, $j_T=j_0 (N/N_0)=j_0(1-T/N_0)$ mentioned above.
The following mean-field picture will clarify further this picture.
We replace the newly fixed spin $s_{T+1}$ by its mean $m^{\rm (eq)}_{T},$ and also use the {}{saddle-point approximation for the integrand in (\ref{eq:IntegZ})
around $x/N = m^{(eq)}_T.$
Then the stochastic rules, (\ref{eq:langevin}) and (\ref{eq:proba}), are replaced by the deterministic rules:}
\beqa \label{eq:compensation}
M_{T+1}&=& M_T+m^{\rm (eq)}_{T},\,
\cr
 m^{\rm (eq)}_{T} 
&=& 
\tanh\inSbracket{ j_0\inRbracket{1-\frac{T}{N_0}} m^{\rm (eq)}_{T}+ \frac{j_0}{N_0}M_T}.
\eeqa 
This recurrence relation 
 tells that $m^{\rm (eq)}_{T+1}=m^{\rm (eq)}_{T}$ The derivation is given in Appendix.\ref{app:MF}.
We would assert that all the above arguments about the (quasi) martingale property hold whether or not the starting state is at the critical point.
\subsection{Statistical ensemble of fixed spins}  
\begin{figure}[t!!]
\centering
\subfigure[\null ]
{\label{fig:evolution-00}
\includegraphics[width=4.2cm,angle=-0.]{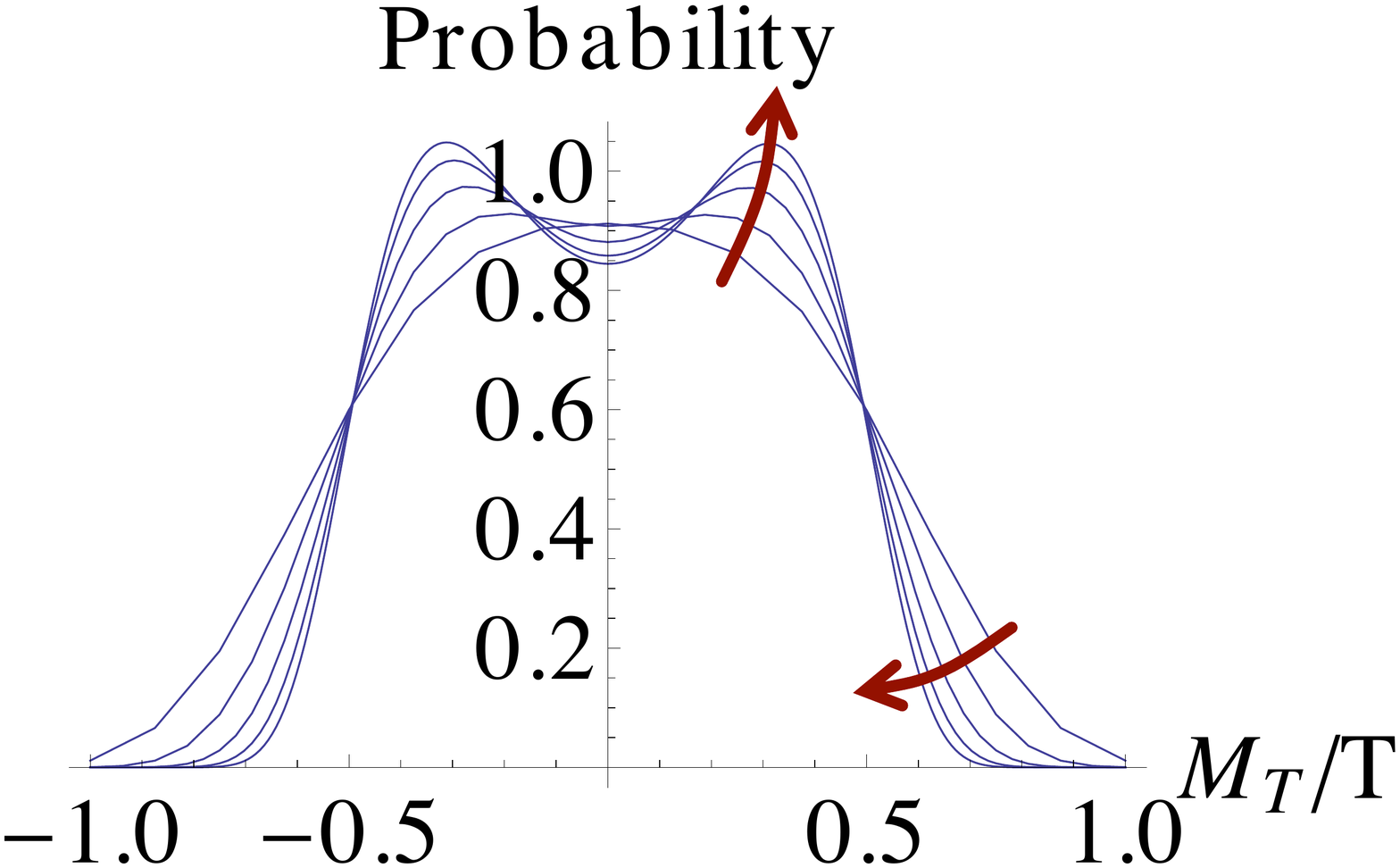}
}  
\hspace{-3mm}
\subfigure[\mbox{ }\null ]
{\label{fig:evolution-11}
\includegraphics[width=4.2cm,angle=-0.]{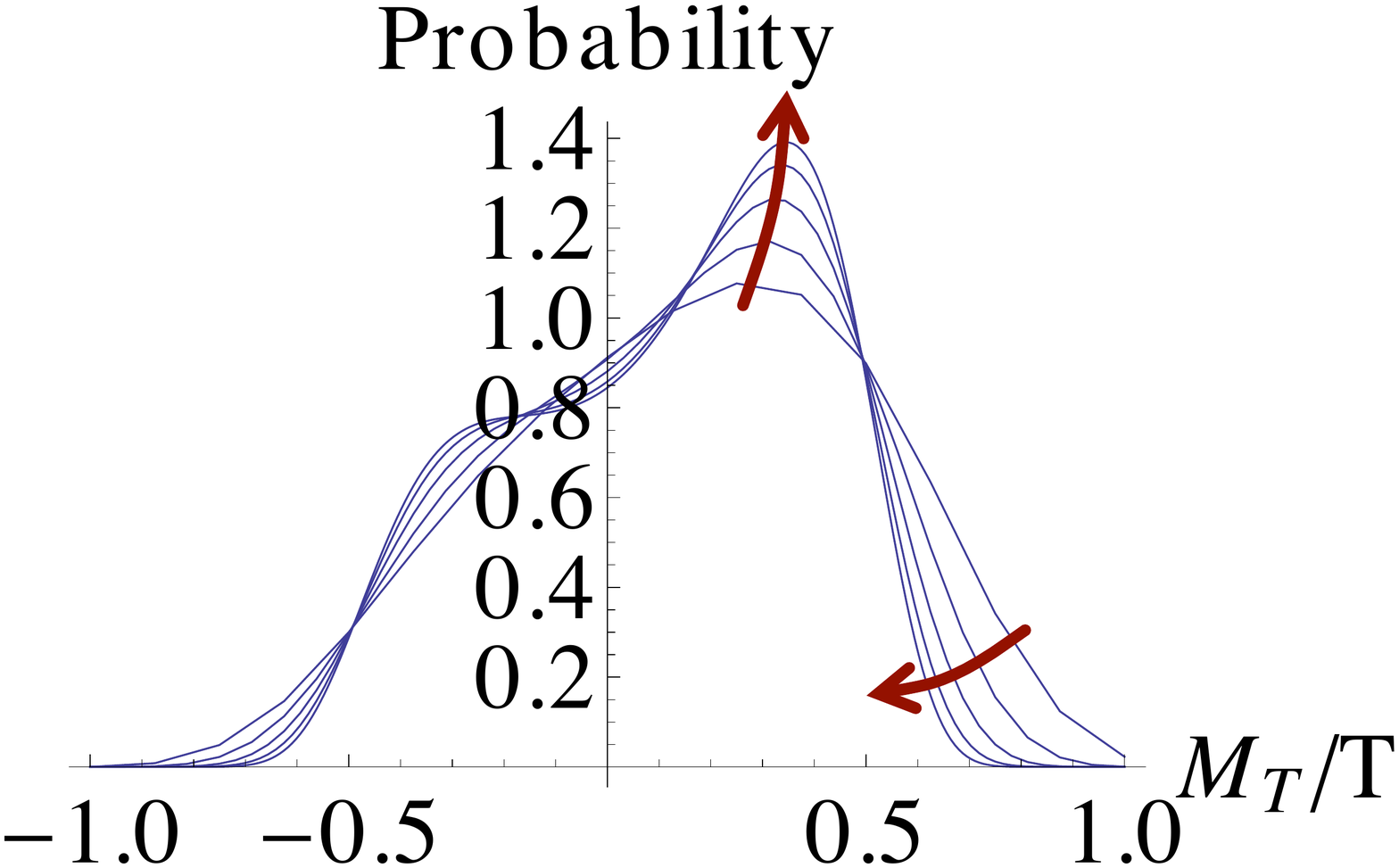}
}  
\hspace{-3mm}
\subfigure[\null ]
{\label{fig:2} 
\includegraphics[width=5.2cm,angle=0.]{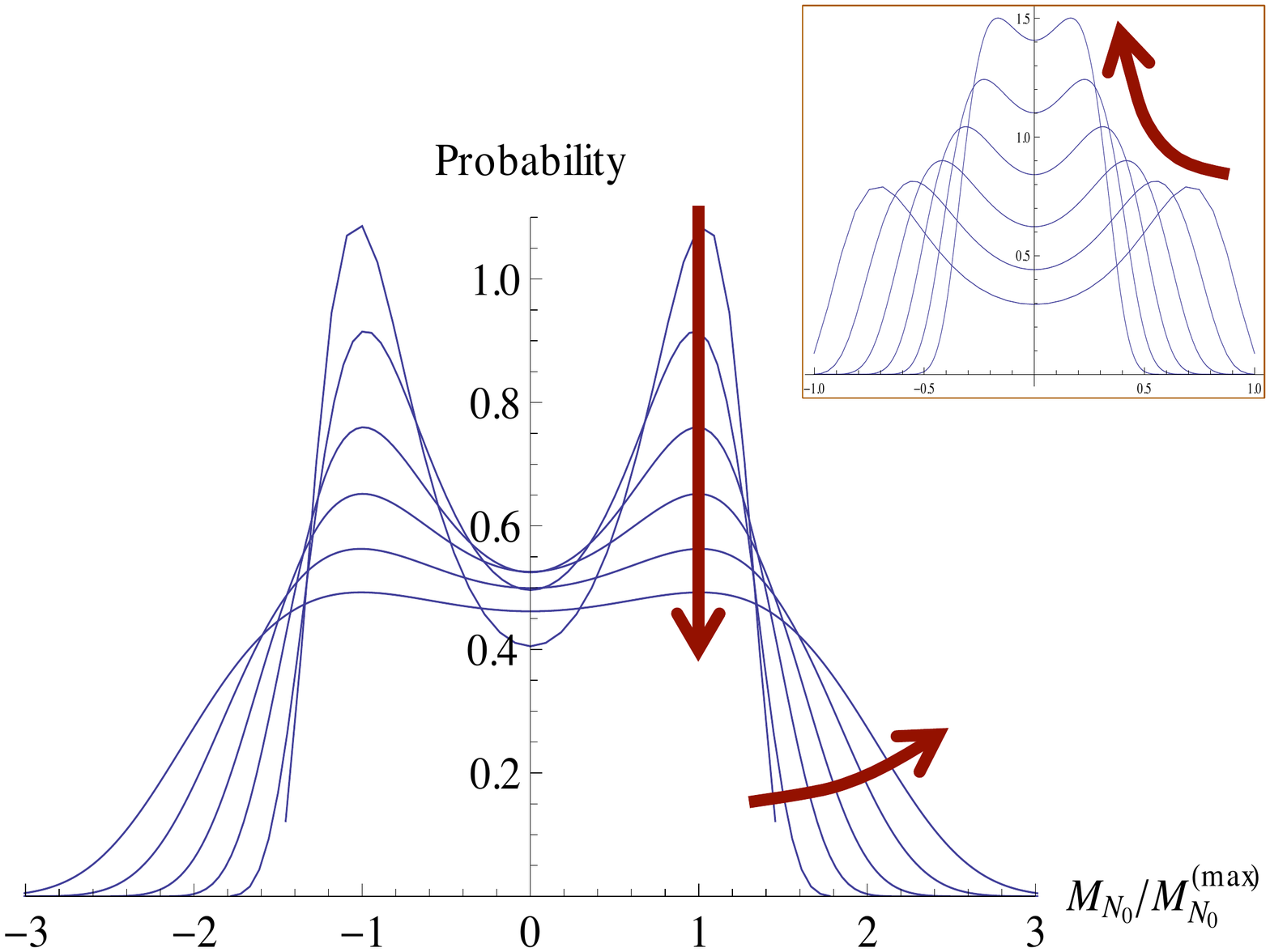}
}
\hspace{-3mm}
\subfigure[]
{\label{fig:1} 
\includegraphics[width=3.2cm,angle=0.]{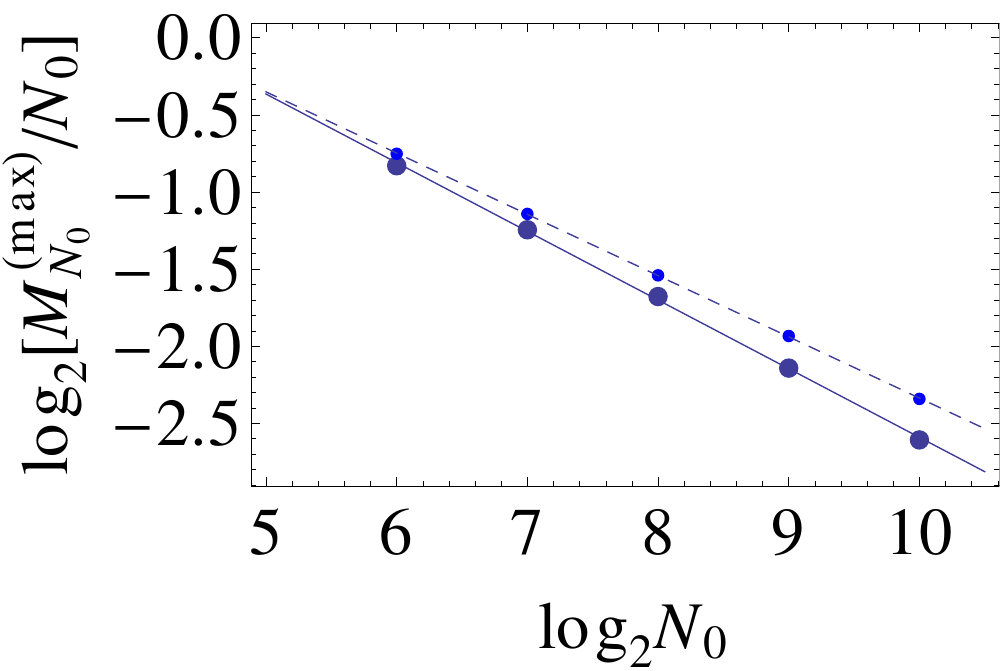}
}
\caption{ \label{fig:evolution}
(a) and (b) Probability distributions of the mean spin value, $M_T/T$, in the quenched part at different stages,
$T=2^k$ with integers $k=4-8$ with the fixed system size, $N_0=2^8=256.$
The initial conditions are (a) $M_0=0$ and (b) $M_1=1$, respectively. 
In both (a) and (b) the increment of $T$ is indicated by the thick red arrows.
(c) Probability density of the magnetization in the final state, $M_{N_0},$ normalized 
 by the maximizer of the probability, {$M_{N_0}/M^{\rm (max)}_{N_0}.$} The system size, $N_0=2^k,$ is varied with integers $k=$5-10. 
{}{The increment in size, $N_0,$ is indicated by the thick red arrows.}
{\it Inset}: The same data but as function of $M_{N_0}{}{/N_0}.$ 
{}{The increment in size, $N_0,$ is indicated by the thick red arrow.}
(d)
Log-log plot of {$M^{\rm (max)}_{N_0} /N_0$ {\it versus} $N_0$} (thick dots). 
The solid line is a linear fitting (i.e., $M^{\rm (max)}_{N_0}/N_0\sim {N_0}^{\alpha}$ ) with the slope {$\alpha\simeq -0.45$}, that is, $M^{\rm (max)}_{N_0}\sim N_0^{1-0.45}.$ Thin dots with the dashed line (the fitted slope {$\alpha\simeq -0.40$})  show the similar results under the condition that the first fixed spin is $s_1=+1$ (see Fig.~\ref{fig:evolution-11}). }
\end{figure}

Because of the memory carried by the martingale property of individual histories, 
we expect an important influence of the initial stochastic process on the 
 subsequent process.
 Fig.~\ref{fig:evolution-00} shows the evolution of the probability distribution of the mean fixed spin value, $M_T/T,$ 
from $T=2^4$ up to $T=N_0=2^8.$ Calculation is done by solving the discrete-``time'' master equation for the biased random walk explained in Fig.~\ref{fig:TN-schema}. 
The progressive quenching causes apparently the splitting of peak in the distribution of $M_T/T.$ The splitting, however, does not mean any bifurcation {}{in the midway} as is evident from the sample histories in Fig.~\ref{fig:m-contour}. The origin of splitting is clear if we plot the conditional probability densities of $M_T/T$ for those histories starting from ${}{M_1=}+1$ (Fig.~\ref{fig:evolution-11}).
(The distribution in Fig.~\ref{fig:evolution-00} can be recovered by taking the average of the result in Fig.~\ref{fig:evolution-11} and its mirror image about the vertical axis.)
In Fig.~\ref{fig:evolution-11} the peak is well off the vertical axis from the beginning and it {}{only} sharpens with the progression of quenching.  {}{These results shows the importance of the initial stochastic events upon the whole history.}
%
\subsection{System size dependence}  
In the inset of Fig.~\ref{fig:2} we show the probability densities of the final quenched {mean} spin, $M_{N_0}/{N_0},$ for the different system sizes, $N_0=2^5$-$2^{10}.$  
We observe that, the larger is the system, the less important is the stochasticity of the initial regime.
If we extrapolate our result to the limit of $N_0\to \infty,$ the distribution of $M_{N_0}/N_0$ will converge to a $\delta$-distribution.
{}{However, within the range of system size we studied, the asymptotic system-size scaling behavior is not observed. 
By the asymptotic system-size scaling we mean
\eqn{ \label{eq:scaling-func}
{\rm Prob}\inRbracket{{\frac{M_{N_0}}{N_0} }}\simeq 
\inv{{N_0}^{\alpha}}\,\,\Psi\,\inRbracket{{\frac{M_{N_0}}{{N_0}^{1+\alpha} } }},
}
with a scaling function $\Psi(\cdot)$ with some exponent $\alpha$. 
To show that (\ref{eq:scaling-func}) is {\it not} the case, we plotted the distribution of the final quenched magnetization
$M_{N_0}$ scaled by the mode value (the value of $M_{N_0}$ for which the probability density takes local maximum) denoted by $M^{\rm (max)}_{N_0}.$ 
The result (Fig.~\ref{fig:2}) shows that the tail of the distribution fattens systematically with the system size, $N_0.$ If the size-scaling (\ref{eq:scaling-func}) were to hold, the mode and the tail should fit to the same scaling function.
Our result shows the transient power law, $M_{N_0}^{\rm (max)}\sim {N_0}^{1+\alpha_{\rm max}}$ with $\alpha_{\rm max}\simeq -0.45$ (see Fig.~\ref{fig:1}, the thick dots) and the fattening of the tails in Fig.~\ref{fig:2} means that the width $\Delta M_{N_0}$ increases more rapidly than $M_{N_0}^{\rm (max)}\sim {N_0}^{0.55}.$ 
For the free spins ($j_0=0$) the width $\Delta M_{N_0}$ should scale diffusively, i.e., $\propto 
{N_0}^{1/2}$ and the above super-diffusive result, together with the unattainability of the asymptotic system-size scaling, reflect the long memory associated to the martingale process. 
The power-law in Fig.~\ref{fig:1} is only transient, not asymptotic one: 
Indeed, if we study $M^{\rm (max)}_{N_0}$ under the initial condition of $M_{T=1}=1$ (see Fig.~\ref{fig:evolution-11}), {}{instead of $M_{T=0}=0$,} the apparent exponent is {$\simeq -0.40,$} unlike $\alpha_{\rm max}$, which is shown by the thin dots with dashed line in Fig.~\ref{fig:1}.
}

\section{{}{Discussion:} Progressive quenching in general context}   
Not being limited to the Ising systems, we expect that the compensation mechanism is a generic feature of the progressive quenching of globally coupled elements, where 
the increment of the field exerted by the quenched elements {}{can largely} compensate the weakening of the internal global coupling among the {}{ unquenched} elements. 
{}{
The emergence of the martingale property indicates that the progressive quenching is the ``neutral operation" that minimally disturbs the unquenched system although the operation is a far from equilibrium operation.}

{}{Martingale property { (\ref{eq:meqISmartingale})
 and its derivation  (\ref{eq:mTexpansion}) are  valid for any ``time'' $T$ except in the final regime ($N\equiv N_0-T\ll N_0$). 
The deviation from (\ref{eq:meqISmartingale}) in the final regime has, however, little consequence on the value of $M_T/T,$ because the latter has already almost converged to $M_{N_0}/N_0.$

The martingale property features the long-term memory in the process and, therefore, the
 importance of the initial regime, $T\ll N_0$. 
In the initial regime, if we use {\it wrongly} the saddle-point approximation to calculate the partition function instead of the full formula (\ref{eq:IntegZ}), the response of $m_T^{\rm (eq)}$ is overestimated
because the mean-field susceptibility diverges at the critical point while the true value should remain of 
$\mathcal{O}(N_0).$ With such saddle-point approximation the size-scaling plot like Fig.\ref{fig:2} 
shows a (fake) convergence to a double peaked scaling function with an exponent, $\alpha\simeq {-0.26}$ (data not shown).  
Also} the continuous path-integral methods such as that of Freidlin-Wentzell \cite{Freidlin-Wentzell} {}cannot be used alone but should be combined with some other techniques because the discreteness of ``time'' $T$ is essential in the initial regime.

{To see the implication of the martingale property, (\ref{eq:meqISmartingale}), let us suppose that the process up to some early time $T$ has been specified. Then we can 
find the value of $M_T$ from the final statistical data of frozen magnetization, $M_{N_0}.$
The first procedure is to calculate the value of $\meq_{N_0}$ that corresponds to each data of $M_T.$ Then their mean value is found to be 
$E[\meq_{N_0}|\{s_1,\ldots,s_T\}] =\meq_T+\mathcal{O}(N_0^{-1})$, which follows from  
the iterative application of  (\ref{eq:meqISmartingale}) \footnote{See the last paragraph of Sec.\ref{subsec:martingale}.} and the ignorance of the small error in the very final steps, as mentioned above. Finally $M_T$ can be inversely calculated from the value of $\meq_T.$
}

{}{With globally coupled models there remains to examine the known consequences of the martingale property as has been done for the fluctuation theorem \cite{Roldan-prX2017}. 
}Progressive quenching of  systems {}{with quenched disorder} is also an open problem.\\
\ind {}{Leaving aside the stochastic process, recent radial Hele-Shaw experiments of two miscible fluids have shown the maintenance of the memory of the initial process. There, the initial pattern is generated by the unstable viscus-fingering \cite{HS-proportionate}, then it evolves later in a self-affine manner (called ``proportionate growth"). We might ask if any compensation mechanism is at work like the one discussed in our study.}
Finally the notion of the (approximate) martingale property of the unquenched part applies also  
to the quenching of the phason field mentioned in the introductory part \cite{phason-freezing-PhA}: The equilibrium expectation value of the field in the unquenched region, $x>Vt,$ is the value of the quenching front at $x=Vt,$ although the variance is divergent in that model.

\begin{acknowledgments}
The authors thank Damien Vandembroucq for having shown their unpublished results. We also thank Luca Peliti and Anton Zadrin for constructive comments. KS also thanks Itamar Procaccia for the valuable comments. {}{We  thank also Edgar Rold\'an for having interest and comments on our preprint.}
BV was financially supported by Universit\'e Paris-Saclay for 2016-2017.
\end{acknowledgments}
\appendix
\section{Martingale property of the Radon-Nikodym derivative 
 and its outcome.}\label{app:RN}
The path probability given in \null{(5
)} in the main text is written as 
\[\mbox{Prob}(\{s_1,\ldots,s_{T}\})=\mbox{Prob}_0(\{s_1,\ldots,s_{T}\})\times 
\inRbracket{\frac{d{\sf P}}{d{\sf P}_0}}_{T},\]
 where $\mbox{Prob}_0(\{s_1,\ldots,s_{T}\})\equiv 2^{-T}$ is the 
probability of unbiased spins (or, equivalently, the unbiased random walk, 
$\{M_0,M_1,\ldots,M_T\}$) and 
\eqn{\label{eq:RN}
\inRbracket{\frac{d{\sf P}}{d{\sf P}_0}}_{T}=
\prod_{T'=1}^{T}\inRbracket{1+s_{T'}m^{\rm (eq)}_{N_0-T'+1,{}\,h_{T'-1}}},
}
which is called the Radon-Nikodym derivative, 
gives the conversion factor from the unbiased random walk to the biased random walk defined by \null{(3
) and (4
)} in the main text. $(d{\sf P}/d{\sf P}_0)_T$ is denoted as $R_T$ in the main text,  but we follow here the convention \cite{Baxter-Rennie1996}.
Then the partial normalization condition, $\sum_{s_{T+1}=\pm 1}\mbox{Prob}(\{s_1,\ldots,s_{T+1}\})=\mbox{Prob}(\{s_1,\ldots,s_{T}\}),$ for $1\le T< N_0$ can be rewritten as 
\beq \label{eq:martingale}
 E_0\inSbracket{\left.\inRbracket{\frac{d{\sf P}}{d{\sf P}_0}}_{T+1} 
 \right| \{s_1,\ldots,s_T\} }
 =\inRbracket{\frac{d{\sf P}}{d{\sf P}_0}}_{T},
\eeq
where $E_0$ means to take the conditional expectation of $(d{\sf P}/d{\sf P}_0)_{T+1}$ over the {\it unbiased} spin (here $s_{T+1}$ only) under the given $\{s_1,\ldots, s_T\}$.
Therefore, the stochastic process, $\{(d{\sf P}/d{\sf P}_0)_{T}\}$ is martingale with respect to the unbiased stochastic process, $\{s_1,\ldots,s_T\}.$ 
Especially, the whole path normalization condition, $\sum_{s_1,\ldots,s_T}\mbox{Prob}(\{s_1,\ldots,s_T\})=1,$ is written in terms of the unconditional expectation over all the unbiased stochastic processes,
\beq \label{eq:OST}
E_0\left[ \inRbracket{\frac{d{\sf P}}{d{\sf P}_0}}_{T}\right]=1.
\eeq
As is anticipated from (\ref{eq:martingale}), the last relation can be obtained by iteratively applying
(\ref{eq:martingale}) down to $T=0,$ when $\inRbracket{\frac{d{\sf P}}{d{\sf P}_0}}_{T}$ is formally unity. This is a general consequence of the martingale property under certain conditions and is
so-called optional stopping theorem (see \cite{Doob1971}) for a review by the founder).

The martingale property of the Radon-Nikodym derivative has recently been brought into physics by \cite{martingale-Gupta2011,Roldan-prX2017},   where $\sf P$ and $\sf P_0$ were the path probabilities for the forward and time-reversed processes, respectively, and (\ref{eq:OST}) was essentially the equalities of Jarzynski \cite{Jarzynski97} and Crooks \cite{JeqCrooks98}.
\vspace{2cm}

\section{Proof of martingale property of $m^{\rm (eq)}_{T}$}\label{app:proof}
The equilibrium spin of the unquenched part, $\meq,$ is the function of the number of unquenched spins, $N,$ and the external field on the unquenched spins, $h,$ which we represent as $\meq_{N,h}.$ Note that the coupling between spins, $j_0/N_0,$ is always fixed during the progressive quenching. Below the description is somehow redundant so as to be clear enough. \\

The equilibrium spin after $T$ spins have been quenched is $\meq_{N_0-T,\frac{j_0}{N_0}M_{T}}.$ We introduce the notation, $m^{\rm (eq)}_{T}\equiv \meq_{N_0-T,\frac{j_0}{N_0}M_{T}}.$
As function of the stochastic process, $\{s_1,\ldots,s_{N_0}\}$, or equivalently, $\{M_0,M_1,\ldots,M_{N_0}\}$ (with $M_0=0$),  
the series, $\{m^{\rm (eq)}_1,\ldots m^{\rm (eq)}_{N_0}\}$ also constitutes a stochastic process. 
We will show that, for $N_0\gg 1$ and $N\equiv N_0-T\sim N_0,$ there holds
\beq \label{eq:Mg-statement-SI} 
E[m^{\rm (eq)}_{T+1}|\{s_1,\ldots, s_{T}\} ]=m^{\rm (eq)}_{T}+\mathcal{O}(\inv{{N_0}^2}).
\eeq
Note that $m^{\rm (eq)}_{T}$ is known when $\{s_1,\ldots, s_{T}\}$ is specified. 
(\ref{eq:Mg-statement-SI}) means that up to the small error of $\mathcal{O}({{N_0}^{-2}}),$  
the stochastic process, $m^{\rm (eq)}_{T},$ is martingale with respect to the process, 
$\{s_1,\ldots,s_{T}\}$, or $\{M_0,M_1,\ldots,M_{T}\}.$ \\

\noi {\it Demonstration:}\\
First we explain the formula of the conditional expectation of $E[m^{\rm (eq)}_{T+1}|\{s_1,\ldots, s_{T}\} ]$ in   (\ref{eq:mTexpansion}) in the main text, that is,
\beq 
E[m^{\rm (eq)}_{T+1}|\{s_1,\ldots, s_{T}\} ]
=\sum_{s_{T+1}=\pm 1} \frac{1+s_{T+1} m^{\rm (eq)}_{T}}{2} \,   
    \meq_{N_0-(T+1),\frac{j_0}{N_0}(M_T+s_{T+1})}.
\eeq    
\ben
\item If $T$ spins have been quenched, we know the values of $\{s_1,\ldots, s_{T}\}$.
Therefore we know also the quenched magnetization, $M_T,$ or the field due to this
magnetization, $h_T=\frac{j_0}{N_0}M_T.$ This, in turn fixes 
the mean equilibrium spin of the unquenched group, 
$m^{\rm (eq)}_{T}=\meq_{N_0-T,\frac{j_0}{N_0}M_T}.$
\item
In order to find the conditional expectation of 
the mean equilibrium spin at stage ${T+1},$ i.e., $m^{\rm (eq)}_{T+1},$ 
we must first know the realization of $m^{\rm (eq)}_{T+1}=M_T+\hat{s}_{T+1}.$ 
Here $\hat{s}_{T+1}$ realizes the value $\pm 1$ with the probability, $(1+s_{T+1} m^{\rm (eq)}_{T})/2,$ respectively. 
Given the value of $ \hat{s}_{T+1}$, which we write as $s_{T+1}$, the mean equilibrium spin at stage ${T+1}$ is $\meq_{N_0-(T+1),\frac{j_0}{N_0}(M_T+s_{T+1})}$.
Therefore, the conditional average of $m^{\rm (eq)}_{T+1}$  under the given  $\{s_1,\ldots, s_{T}\}$ reads
as the formula above.
\een
Next we do the sum over $s_{T+1}$:
\begin{widetext}
{
\beqa
 E[m^{\rm (eq)}_{T+1}|\{s_1,\ldots, s_{T}\} ]
&=&
\sum_{s_{T+1}=\pm 1} \frac{1+s_{T+1} m^{\rm (eq)}_{T}}{2} \,   
    \meq_{N_0-(T+1),\frac{j_0}{N_0}(M_T+s_{T+1})}
\cr
&&= 
\inv{2}\inSbracket{\meq_{N_0-(T+1),\frac{j_0}{N_0}(M_T+1)}+\meq_{N_0-(T+1),\frac{j_0}{N_0}(M_T-1)}}
+
\frac{m^{\rm (eq)}_{T}}{2}\inSbracket{\meq_{N_0-(T+1),\frac{j_0}{N_0}(M_T+1)}-\meq_{N_0-(T+1),\frac{j_0}{N_0}(M_T-1)}}. \nonumber
\eeqa
}
\end{widetext}
Then on the r.h.s. we expand $m^{\rm (eq)}_{N_0-(T+1),\frac{j_0}{N_0}(M_T+s_{T+1})}$ 
around $m^{\rm (eq)}_{T}=\meq_{N_0-T,\frac{j_0}{N_0}M_T}$
with ignoring the terms of $\mathcal{O}({N_0}^{-2})$ such as   $\partial^2 \meq_{N,h}/\partial N^2.$ 
The result is 
\beq \label{eq:testMg}
E[m^{\rm (eq)}_{T+1}|\{s_1,\ldots, s_{T}\} ]
=m^{\rm (eq)}_{T}-\frac{\partial m^{\rm (eq)}_{T}}{\partial N}+ \frac{j_0}{N_0}m^{\rm (eq)}_{T}\frac{\partial m^{\rm (eq)}_{T}}{\partial h} +
\mathcal{O}({N_0}^{-2}),
\eeq
where $m^{\rm (eq)}_{T}=\meq_{N_0-T,h_T}$ is regarded as function of 
 $N=N_0-T$ and $h_T=\frac{j_0}{N_0}M_T$. 
Even though we are tempted to use the mean-field approximation, the dependence of $\meq$ on the size, $N,$ does not allow this. 
We will see that the second and the third terms cancel each other to $\mathcal{O}({N_0}^{-1}).$  
To estimate $\frac{\partial m^{\rm (eq)}_{T}}{\partial N}$ we introduce the canonical partition function
for the $N$ (unquenched) spins under the field $h$:
\[Z_{N,h}=\sum\cdots \sum_{s_1,\ldots,s_N} e^{N[j_T\frac{
{\overline{m}}^2}{2}\,
+h {\overline{m}}]},
\]
where we have introduced $ j_{T}\equiv \frac{N}{N_0}j_0=\inRbracket{1-\frac{T}{N_0}}j_0,$ and 
${\overline{m}}\equiv \sum_{k=1}^N s_k/N$ is the empirical mean of the unquenched spins.
Since $N$ is the total number of {\it unquenched} spins, the notation $\overline{m}$ will be justifiable. 
Note that, as a system of $N$ unquenched spins, the role of the coupling parameter $j_0$ in the {\it total} energy function 
of $N_0$ spins is played by the effective one, $j_{T} (<j_0).$ 
Using the partition function $Z_{N,h}$ the mean spin is given by the canonical average,
$\meq_{N,h}= + {N^{-1}} {\partial \ln Z_{N,h}}/{\partial h}.$
The direct estimation of $Z_{N,h}$ with $N\gg 1$ gives $Z_{N,h}\sim e^{N\phi(j_{T} ,h)}$ 
with 
\begin{widetext}
\beq
\phi(j_{T},h)=\inSbracket{
j_{T}\frac{\overline{m}^2}{2}+h \overline{m} 
-\frac{1+ \overline{m}}{2}\ln\frac{1+ \overline{m}}{2}
-\frac{1- \overline{m}}{2}\ln\frac{1 -\overline{m}}{2}}_{\overline{m}=m^{\rm (eq)}_{N,h}}
+\inv{2N}\log(1-j_T (1-\tanh(j_T m^{\rm (eq)}_{N,h}+h))),
\eeq
\end{widetext}
where we retained the {\it off-shell} expression in the square bracket 
so that the expression allows to take the derivatives of $\phi$ and 
the condition $\partial\phi/\partial \overline{m}=\mathcal{O}(N^{-1})$ applied to this expression
leads to ${\overline{m}=m^{\rm (eq)}_{N,h}}.$
We then have $\meq_{N,h} =  {\partial \phi(j_{T} ,h)}/{\partial h} +\mathcal{O}({N_0}^{-1}),$
and the only $N$-dependence of ${\meq}_T$ comes through $j_{T}$ in $\phi(j_T,h),$ that is,
\beqa
\frac{\partial m^{\rm (eq)}_{T}}{\partial N} 
&=& \frac{\partial j_{T}}{\partial N}\times \frac{\partial \meq_{N,h}}{\partial j_{T}}
=\frac{j_0}{N_0}\times \frac{\partial}{\partial h }\inRbracket{\inv{N} \frac{\partial \ln Z_{N,h}}{\partial j_{T}}
}
\cr &=& \frac{j_0}{N_0}\times \frac{\partial}{\partial h}\inAverage{\frac{ {\overline{m}}^2}{2}}^{\!\!\rm (eq)},
\eeqa
where the terms of $\mathcal{O}({N_0}^{-2})$ are ignored and 
$\inAverage{\cdot}^{\rm (eq)}$ means the canonical equilibrium average.
Substituting the last result for ${\partial m^{\rm (eq)}_{T}}/{\partial N}$ in (\ref{eq:testMg}),
we arrive (note that $m^{\rm (eq)}_{T}=\inAverage{ \overline{m}}^{\rm (eq)}$)
\beqa \label{eqapp:testMg}
&& E[m^{\rm (eq)}_{T+1}|\{s_1,\ldots, s_{T}\} ]
\cr &&=m^{\rm (eq)}_{T}+   \frac{j_0}{2N_0}\frac{\partial }{\partial h}
\inSbracket{\inAverage{ \overline{m}^2}^{\rm (eq)}-\inRbracket{\inAverage{ \overline{m}}^{\rm (eq)}}^2}
 +
\mathcal{O}({N_0}^{-2}).
\eeqa
As is clear from the large deviation form of $Z_{N,h}$ mentioned above the equilibrium variance of $\overline{m}$
is  $[\inAverage{ \overline{m}^2}^{\rm (eq)}-(\inAverage{ \overline{m}}^{\rm (eq)})^2]=\mathcal{O}({N}^{-1})$  and is, therefore, $\mathcal{O}({N_0}^{-1})$ { unless} $N\ll N_0$.
We thus arrive at the martingale property 
(\ref{eq:Mg-statement-SI}) up to the error of $\mathcal{O}({N_0}^{-2}).$
{Notice that $E[s_{T+1}|\{s_1,\ldots,s_T\}]=m^{\rm (eq)}_{T}$ is a definition of our model
and is different from  (\ref{eq:Mg-statement-SI}). Notice also that 
(\ref{eq:Mg-statement-SI}) does {\it not} imply $E[s_{T+1}|\{s_1,\ldots,s_T\}]\stackrel{!}{=} s_{T}+\mathcal{O}({N_0}^{-2}).$}
 That the error is  $\mathcal{O}({N_0}^{-2})$ not $\mathcal{O}({N_0}^{-1})$ is crucial as stressed in the main text.

\section{Mean-field argument of the mechanism behind the martingale property}\label{app:MF}
In the mean-field argument, we replace the stochastically quenched spin, $\hat{s}_{T+1},$
by its statistical mean, $m^{\rm (eq)}_{T}\equiv \inAverage{\hat{s}_{T+1}}^{\rm (eq)},$ given the already
quenched spin, $M_T=\sum_{k=1}^T s_k.$
We show the following statement: \\

\noi
The sequence $\{m^{\rm (eq)}_{T}\}$ generated by the following set of equations with $1\le T<N_0$ 
can satisfy $m^{\rm (eq)}_{T+1}=m^{\rm (eq)}_{T}.$
\beq \label{eq:MTstep-1}
M_{T+1}=M_T+m^{\rm (eq)}_{T}
\eeq
\beq \label{eq:mTcond-1}
m^{\rm (eq)}_{T}=\tanh\inSbracket{j_0\inRbracket{1-\frac{T}{N_0}}m^{\rm (eq)}_{T}+\frac{j_0}{N_0}M_T},
\eeq

\mbox{}\\
The proof does not require the explicit solution of the above implicit equation.
First we rewrite the r.h.s. of (\ref{eq:mTcond-1}) to have
\[
m^{\rm (eq)}_{T}=\tanh\inSbracket{j_0\inRbracket{1-\frac{T+1}{N_0}}m^{\rm (eq)}_{T}+\frac{j_0}{N_0}(M_T+m^{\rm (eq)}_{T})}.
\]
For $M_T+m^{\rm (eq)}_{T}$ we substitute (\ref{eq:MTstep-1}) to have
\[
m^{\rm (eq)}_{T}=\tanh\inSbracket{j_0\inRbracket{1-\frac{T+1}{N_0}}m^{\rm (eq)}_{T}+\frac{j_0}{N_0}M_{T+1}}.
\]
On the other hand, if we apply (\ref{eq:mTcond-1}) for $T\mapsto T+1,$ we have
\[
m^{\rm (eq)}_{T+1}=\tanh\inSbracket{j_0\inRbracket{1-\frac{T+1}{N_0}}m^{\rm (eq)}_{T+1}+\frac{j_0}{N_0}M_{T+1}}.
\]
Therefore, the above generating rule allows $m^{\rm (eq)}_{T+1}=m^{\rm (eq)}_{T}$ by properly choosing the branch of solutions at each step.
Thus the weakening of the effective spin-spin coupling, 
$j_{\rm eff}=j_0\inRbracket{1-\frac{T}{N_0}},$ is exactly compensated by
the increment of the amplitude of the quenched field, $|h_{\rm T}|=
\frac{j_0}{N_0}|M_T|$ and, in consequence,
the mean equilibrium magnetization of the unquenched spins, $|m^{\rm (eq)}_{T}|,$ is maintained stationary.

\bibliographystyle{apsrev4-1.bst}    \bibliography{ken_LNP_sar}
\end{document}